\documentclass[preprint,aps,prd,showpacs,floatfix,nofootinbib]{revtex4}
\usepackage{graphicx}
\usepackage{amsmath}
\usepackage{amssymb}
\usepackage{bm}
\usepackage{bbm}
\begin{document}
\title{\mbox{}\\[10pt]
Exclusive heavy quarkonium $\bm{+}$ $\bm{\gamma}$ production
\\
from $\bm{e^+ e^-}$ annihilation into a virtual photon
}
\author{Hee~Sok~Chung}
\affiliation{Department of Physics, Korea University, Seoul 136-701, Korea}
\author{Jungil~Lee}
\affiliation{Department of Physics, Korea University, Seoul 136-701, Korea}
\author{Chaehyun~Yu}
\affiliation{Department of Physics, Korea University, Seoul 136-701, Korea}

\begin{abstract}
We compute the cross section for exclusive production of a photon
associated with a heavy quarkonium $H$ of charge-conjugation parity $C=+1$
from $e^+e^-$ annihilation into a virtual photon 
at the center-of-momentum energy $\sqrt{s}=10.58$~GeV.
The nonrelativistic QCD factorization formulas for the differential 
and total cross sections
are obtained at leading order in the strong coupling and in the relative
velocity of the heavy quark in the quarkonium rest frame.
The predicted cross sections for the $S$-wave spin-singlet cases are
about $80$~fb and $50$~fb for $H=\eta_c$ and $\eta_c(2S)$, respectively.
Among $P$-wave spin-triplet charmonia, $\chi_{c1}$ has a particularly
large cross section of about $14$~fb. The cross sections for bottomonium
states $\eta_b$ and $\chi_{bJ}$ are about $3$~fb.
A rough estimate of the background reveals that the signal significances
for charmonium processes are sufficiently large enough to be
detected with ease with the integrated luminosities available at
the present $B$ factories.
\end{abstract}



\maketitle


\section{Introduction 
\label{sec:intro}}
The measurements of the cross section for exclusive $J/\psi+\eta_c$
production from $e^+e^-$ annihilation carried out by the
Belle~\cite{Abe:2002rb,Abe:2004ww} and BABAR~\cite{Aubert:2005tj}
collaborations have triggered rapid progress in the heavy-quarkonium
theory based on the nonrelativistic QCD (NRQCD) factorization
approach~\cite{Bodwin:1994jh}. The original empirical cross section
in Ref.~\cite{Abe:2002rb} was greater than the theoretical predictions
by Braaten and Lee~\cite{Braaten:2002fi} and by Liu, He, and
Chao~\cite{Liu:2002wq} by an order of magnitude.
Scenarios that the signal may contain other final
states like $J/\psi+J/\psi$~\cite{Bodwin:2002fk,Bodwin:2002kk}
or $J/\psi+$glueball~\cite{Brodsky:2003hv} were ruled out by
an updated Belle analysis~\cite{Abe:2004ww}. There was an argument
that an appropriate choice of the light-cone wave 
function~\cite{Bondar:2004sv} may enhance the theoretical prediction
significantly, which was disfavored by a following
study~\cite{Bodwin:2006dm}. Zhang, Gao, and Chao reported
large corrections of next-to-leading order 
in the strong coupling $\alpha_s$~\cite{Zhang:2005cha} and the result
was confirmed by Gong and Wang~\cite{Gong:2007db}, but the corrections
were not large enough to resolve the discrepancy between experiment
and theory. In the mean time, Bodwin, Kang, and Lee introduced
a new method of resumming relativistic
corrections of a class of color-singlet contributions to all orders 
in $v$~\cite{Bodwin:2006dn}, where $v$ is the relative
velocity of the heavy quark or antiquark in the meson rest frame. 
The method was applied to compute the relativistic corrections
to $e^+e^-\to J/\psi+\eta_c$~\cite{Bodwin:2006ke}.
Finally, Bodwin, Lee, and Yu reported that the discrepancy
had been resolved within errors~\cite{Bodwin:2007ga},
after including both the resummed relativistic corrections
and next-to-leading order QCD corrections by Zhang, Gao, and Chao.
In addition, the first proof of the factorization theorem for
exclusive charmonium production in $B$ decay and 
exclusive two-charmonium production from $e^+e^-$
annihilation to all orders in perturbation theory in QCD was
presented very recently by Bodwin, Garcia i Tormo, and
Lee~\cite{Bodwin:2008nf}.

As shown in Ref.~\cite{Braaten:2002fi}, at leading order in $v$,
a non-negligible amount of the cross section for the exclusive
production of charge-conjugation parity $C=+1$ charmonium ($H$)
associated with a $J/\psi$ from $e^+e^-$ annihilation
into a virtual photon comes from the interference between QCD
and QED subprocesses, where the QED subprocess involves
$\gamma^*\to H+\gamma^*$ followed by the photon fragmentation
$\gamma^*\to J/\psi$. The interference contributions are about 
$+30$\%, $+5$\%, $-6$\%, and $10$\% for $H=\eta_c$, $\chi_{c0}$,
$\chi_{c1}$, and $\chi_{c2}$, respectively~\cite{Braaten:2002fi}.
This means that the pure QED contributions are
about $10$\%, $0.3$\%, $0.4$\%, and $1$\% of the total cross
sections, respectively. We notice that,
in comparison with the QED subprocess for $e^+e^-\to H+J/\psi$,
the rate for $e^+e^-\to H+\gamma$
is enhanced by a scaling factor of $1/(\alpha v^3)\sim 10^3$
for a charmonium, where $\alpha$ is the QED coupling.
Based on this rough estimate, we expect that the cross section
for $e^+e^-\to H+\gamma$ can be greater than that
for $e^+e^-\to H+J/\psi$ by 2 orders of magnitude for
$H=\eta_c$. For the $P$-wave spin-triplet charmonia, the enhancement
factors range from about 1 to 10. Therefore, as long as the severe 
background from $e^+e^-\to X+\gamma$ is well controlled in the recoil
mass ($m_X$) regions near the $H$ resonances, the processes
$e^+e^-\to H+\gamma$ can be detected by analyzing the photon energy
spectrum in $e^+e^-\to X+\gamma$.

In this paper, we compute the cross section for the exclusive
process $e^+ e^- \to H + \gamma$ at the center-of-momentum (CM)
energy $\sqrt{s}=10.58$~GeV,
where $H$ is a heavy quarkonium of charge-conjugation parity $C=+1$.
We consider $S$-wave spin-singlet states $\eta_c$, $\eta_c(2S)$,
and $\eta_b$ and $P$-wave spin-triplet states $\chi_{cJ}$ and 
$\chi_{bJ}$, where $J=0$, $1$, and $2$.\footnote{Throughout this
paper we suppress the identifier $(1P)$ for any
$1P$ quarkonium.}
Perturbative calculation is carried out at order $\alpha^3\alpha_s^0$
and at leading order in $v$. The remainder of this paper is organized
as follows. In Sec.~\ref{sec:pert}, we compute the short-distance
coefficients for the NRQCD factorization formulas for the
$e^+e^-\to H+\gamma$ cross sections. The numerical values for the
input parameters such as long-distance NRQCD matrix elements are
given in Sec.~\ref{sec:input}. The cross sections for
$e^+e^-\to H+\gamma$ at the $B$ factories are predicted in
Sec.~\ref{sec:num}, where a rough estimate on the background is also
given. Then, we summarize our results in Sec.~\ref{sec:summary}.
\section{Perturbative calculation
\label{sec:pert}}
In this section, 
we derive the NRQCD factorization formula for the exclusive process
$e^+e^-\to H+\gamma$ at leading order in both $\alpha_s$ and $v$,
where $H$ is a heavy quarkonium of charge-conjugation parity $C=+1$. 
The quarkonia $H$ that we consider in this paper are the $S$-wave
spin-singlet (${}^1S_0$) state, $\eta_Q$, and the $P$-wave
spin-triplet (${}^3P_J$) states, $\chi_{QJ}$, for $J=0$, 1, and 2,
where the heavy quark $Q$ is the charm or the bottom quark. Because
the $C$ parity of the final state is $-1$, at leading order in
$\alpha$, an exclusive $H+\gamma$ final state can be produced from
$e^+e^-$ annihilation into a virtual photon. A heavy quarkonium with 
$C=-1$ can be produced associated with a photon from $e^+e^-$
annihilation into two photons, which we do not consider here.
\subsection{Amplitude
\label{sec:amplitude}}
\begin{figure}
\centerline{
\includegraphics[height=3.5cm]{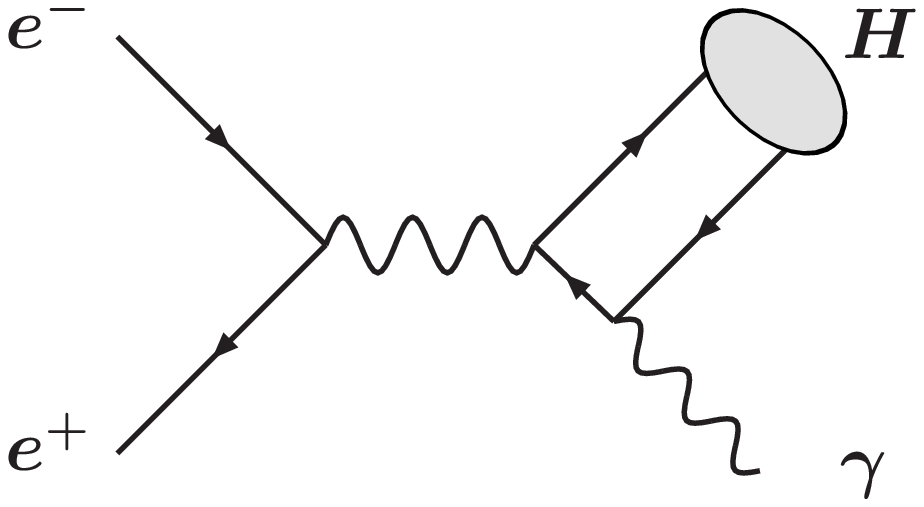}
\quad\quad
\includegraphics[height=3.5cm]{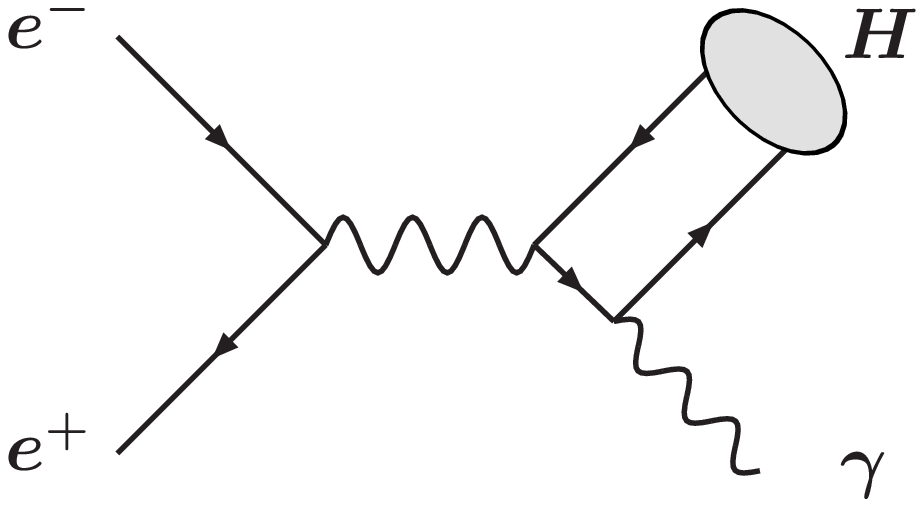}
}
\caption{
Feynman diagrams for the exclusive production of a $C=+1$ heavy
quarkonium associated with a photon from $e^+e^-$ annihilation
into a virtual photon at leading order in $\alpha$ and $\alpha_s$.
}
\label{fig1}
\end{figure}
The Feynman diagrams for the exclusive process
$e^+(k_2)\, e^-(k_1) \to H (P,\lambda_H)+\gamma(k,\lambda)$
at order $\alpha^3\alpha_s^0$
are shown in Fig.~\ref{fig1}. Here, $k_1$, $k_2$, $P$, and $k$
are the momenta for the $e^-$, $e^+$, $H$, and $\gamma$,
respectively. The helicities of the $H$ and $\gamma$ are denoted by
$\lambda_H$ and $\lambda$, respectively. 
The $S$-matrix element for the process
is given by
\begin{equation}
\label{amp0}%
-i \mathcal{M}_H(\lambda_H,\lambda) =
-i \frac{e}{s}
L_\mu\mathcal{A}_H^{\mu\nu}(\lambda_H)\epsilon_{\gamma\nu}^{*}(\lambda),
\end{equation}
where $-e$ is the electric charge of the electron and
$\epsilon_\gamma$ is the polarization four-vector of the photon.
The leptonic current $L_\mu$ in Eq.~(\ref{amp0}) is defined by
\begin{equation}
\label{lepton}%
L_\mu = \bar{v}(k_2) \gamma_\mu u(k_1).
\end{equation}
The factor $\mathcal{A}_H^{\mu\nu}\epsilon_{\gamma\nu}^\ast$
in Eq.~(\ref{amp0}) corresponds to the amplitude for
$\gamma^*(Q)\to H(P,\lambda_H)+\gamma(k,\lambda)$,
where $Q$ is the momentum of the virtual photon $\gamma^*$.
\subsection{Effective vertex
\label{sec:vertex}}
$\mathcal{A}^{\mu\nu}_H$ in Eq.~(\ref{amp0}) contains nonperturbative
contributions of the $Q\bar{Q}$ pair that evolves into the heavy
quarkonium $H$. In the NRQCD factorization approach~\cite{Bodwin:1994jh},
the hadronic tensor $\mathcal{A}^{\mu\nu}_H$ can be expanded as a linear
combination of NRQCD matrix elements that involve the long-distance
nature of the $Q\bar{Q}$ pair. A corresponding short-distance coefficient
of each NRQCD matrix element is insensitive to the long-distance nature of
the pair and calculable perturbatively. According to the velocity
scaling rules of NRQCD, NRQCD matrix elements are
classified in powers of $v$~\cite{Bodwin:1994jh}. Thus,
the expansion can be truncated at a given order of accuracy.
The velocity scaling factor of an NRQCD matrix element is determined by
the spectroscopic state of the $Q\bar{Q}$ pair,
$[Q\bar{Q}]_1({}^{2s+1}L_J)$, and that of the $H$ into which the
$Q\bar{Q}$ pair evolves. Here, $s$, $L$, and $J$ are the spin,
orbital angular momentum, and total angular momentum of the $Q\bar{Q}$
pair, respectively, and the subscript $1$ indicates that
the pair is in a color-singlet state. 

For the exclusive process $\gamma^*\to H+\gamma$,
the color-singlet $Q\bar{Q}$ pair with the spectroscopic state identical
to that of the $H$ contributes at leading order in $v$.
In order to determine the corresponding short-distance coefficient, 
we compute the $Q\bar{Q}$ analog $\mathcal{A}^{\mu\nu}_Q$ of the hadronic
tensor $\mathcal{A}^{\mu\nu}_H$, where 
$\mathcal{A}^{\mu\nu}_Q({}^{2s+1}L_J) \epsilon_{\gamma\nu}^\ast$
is the amplitude for the perturbative process
$\gamma^\ast (Q) \to [Q(p)\, \bar{Q} (\bar{p})]_1({}^{2s+1}L_J)+\gamma(k)$.
Here, the four-momenta $p$ and $\bar{p}$ for the $Q$ and the $\bar{Q}$,
respectively, can be expressed in terms of the total momentum
$P$ and half their relative momentum $q$:
\begin{subequations}
\begin{eqnarray}
p &=& \frac{1}{2}P + q, \\
\bar{p} &=& \frac{1}{2}P - q, 
\end{eqnarray}
\end{subequations}
where $P\cdot q = 0$. In the $Q\bar{Q}$ rest frame, 
$P=(2 E, \bm{0})$, $q=(0, \bm{q})$, 
$E=(m^2+\bm{q}^2)^{1/2}$, and $m$ is the mass of the heavy quark $Q$.
At leading order in $v$,  $P$ can be identified as the quarkonium
momentum. The amplitude for the perturbative process
$\gamma^*(Q)\to Q(p)\bar{Q}(\bar{p})+\gamma(k,\lambda)$ is given by
\begin{equation}
\label{ampA}%
\mathcal{M}^\mu[\gamma^*(Q)\to
 Q(p)\bar{Q}(\bar{p})+\gamma(k,\lambda)]
=
\bar{u}(p)\mathcal{A}^{\mu\nu}v(\bar{p})
\epsilon^*_{\gamma \nu}(\lambda).
\end{equation}
The tensor $\mathcal{A}^{\mu\nu}$ in Eq.~(\ref{ampA}) is defined by
\begin{equation}
\label{AMN}%
\mathcal{A}^{\mu\nu}
=
e_Q^2e^2
\left[
\gamma^\mu \Lambda(-\bar{p}-k)\gamma^\nu
+
\gamma^\nu \Lambda(p+k)\gamma^\mu\right]
\otimes
\mathbbm{1},
\end{equation}
where $e_Q$ is the fractional electric charge of the heavy quark $Q$,
$\Lambda(p)=(\,/\!\!\!p+m)/(p^2-m^2)$, and $\mathbbm{1}$ is the 
unit matrix of the color SU$(N_c)$ for $N_c=3$.

$\mathcal{A}^{\mu\nu}$ in Eq.~(\ref{ampA}) may contain various
$[Q(p)\, \bar{Q} (\bar{p})]_1({}^{2s+1}L_J)$ contributions. 
The $Q\bar{Q}$ analog $\mathcal{A}_Q^{\mu\nu}({}^{2s+1}L_J)$
of the hadronic tensor can be obtained from $\mathcal{A}^{\mu\nu}$ in
Eq.~(\ref{AMN}) by restricting the $Q\bar{Q}$ pair
to have an appropriate spectroscopic state. A
given spin state of the color-singlet $Q\bar{Q}$ pair can be 
projected out from Eq.~(\ref{ampA}) by replacing the outer
product $v(\bar{p}) \bar{u}(p)$ with the corresponding spin projector
given in Ref.~\cite{Bodwin:2002hg}. The color-singlet 
spin-singlet projector is denoted by $\Pi_1$ and the 
color-singlet spin-triplet projector is by
$\Pi_3^\mu \epsilon_{\mu}^\ast(\lambda_s)$, where
$\epsilon(\lambda_s)$ is the polarization for the spin-triplet
state with the helicity $\lambda_s$. Expressions
for the $\Pi_1$ and $\Pi_3^\mu$ are given by~\cite{Bodwin:2002hg}
\begin{subequations}
\begin{eqnarray}
\label{pi1}%
\Pi_1 &=& \frac{
(/\!\!\!\bar{p}-m)
\gamma^5
(\,/\!\!\!\!P\!+\!2E)
(/\!\!\!{p}+m)
}{4\sqrt{2}E(E+m)}
\otimes
\frac{1}{\sqrt{N_c}} \mathbbm{1},
\\
\label{pi3}%
\Pi_3^\mu &=&
-\frac{
(/\!\!\!\bar{p}-m)
\gamma^\mu
(\,/\!\!\!\!P\!+\!2E)
(/\!\!\!{p}+m)
}{4\sqrt{2}E(E+m)}
\otimes
\frac{1}{\sqrt{N_c}} \mathbbm{1}.
\end{eqnarray}
\end{subequations}

We still need to project out an appropriate
orbital-angular-momentum state. In order to project out the $S$-wave
contribution at leading order in $v$, one must put $q=0$. By making
use of Eqs.~(\ref{AMN}) and (\ref{pi1}), we obtain the $Q\bar{Q}$
analog $\mathcal{A}^{\mu\nu}_Q({}^1S_0)$ of the hadronic tensor
for the $S$-wave color-singlet spin-singlet state:
\begin{equation}
\label{amp1s0}%
\mathcal{A}_Q^{\mu\nu}({}^1S_0) =
\textrm{Tr}[
\mathcal{A}^{\mu\nu}
\Pi_1 ]\Big|_{q=0},
\end{equation}
where the trace is over both spin and color indices.

The $P$-wave contribution at leading order in $v$ is proportional
to the first derivative of Eq.~(\ref{ampA})
with respect to $q$.
By making use of Eqs.~(\ref{AMN}) and (\ref{pi3}), 
we obtain the $Q\bar{Q}$ analog
$\epsilon^*_\rho(\lambda_s)%
\epsilon^*_\tau(\lambda_\ell)%
\mathcal{A}_Q^{\mu\nu\rho\tau}$
of the hadronic tensor
for the $P$-wave color-singlet spin-triplet state,
where $\epsilon(\lambda_\ell)$ is the polarization for the $P$-wave
orbital-angular-momentum state and 
\begin{equation}
\label{AQP}%
\mathcal{A}_Q^{\mu\nu\rho\tau}=
\frac{\partial}{\partial q_\rho}
\textrm{Tr}[
\mathcal{A}^{\mu\nu}
\Pi_3^\tau ]\Big|_{q=0}.
\end{equation}
Like Eq.~(\ref{amp1s0}) the trace in Eq.~(\ref{AQP}) is over 
both spin and color indices.
The $Q\bar{Q}$ analog $\mathcal{A}^{\mu\nu}_Q({}^3P_J,\lambda_H)$
of the hadronic tensor 
for the $[Q\bar{Q}]_1({}^3P_J)$
for $J=0$, $1$, and $2$ can be read off by projecting out the diagonal,
antisymmetric, and symmetric traceless components with respect to
the vector indices for $s$ and $L$~\cite{Braaten:2002fi}:
\begin{subequations}
\label{amp3pj}%
\begin{eqnarray}
\label{amp3p0}%
\mathcal{A}_Q^{\mu\nu}({}^3P_0) &=&
\mathcal{A}_Q^{\mu\nu\rho\tau} \frac{1}{\sqrt{3}}I_{\rho\tau},\\
\label{amp3p1}%
\mathcal{A}_Q^{\mu\nu}({}^3P_1,\lambda_H) &=&
\mathcal{A}_Q^{\mu\nu\rho\tau} 
\frac{i}{2E\sqrt{2}}\epsilon_{\rho\tau\alpha\beta} P^{\alpha}
\epsilon_H^{\ast\beta}(\lambda_H), \\
\label{amp3p2}%
\mathcal{A}_Q^{\mu\nu}({}^3P_2,\lambda_H) &=&
\mathcal{A}_Q^{\mu\nu\rho\tau}
\epsilon_{H \rho\tau}^{\ast}(\lambda_H),
\end{eqnarray}
\end{subequations}
where $\lambda_H$ is the helicity of $H$.
$\epsilon_H^\beta$ in Eq.~(\ref{amp3p1}) and 
$\epsilon_H^{\alpha\beta}$ in Eq.~(\ref{amp3p2})
are the spin-1 polarization vector for the $[Q\bar{Q}]_1({}^3P_1)$ pair and
the spin-2 polarization tensor for the $[Q\bar{Q}]_1({}^3P_2)$ pair, 
respectively. 
Note that $\epsilon_H^{\alpha\beta}=\epsilon_H^{\beta\alpha}$
and ${{\epsilon_H}^{\alpha}}_{\alpha}=0$.
Because we consider the color-singlet contributions
at leading order in $v$, the polarization vector and tensor are
treated to be identical to those for $H({}^3P_1)$ and $H({}^3P_2)$,
respectively. The tensor $I_{\mu\nu}$ appearing in Eq.~(\ref{amp3pj})
is defined by
\begin{equation}
I_{\mu\nu} = - g_{\mu\nu} + \frac{P_\mu P_\nu}{(2 E)^2}.
\end{equation}

As a result, we obtain the effective vertex for
the perturbative process
$\gamma^*\to [Q\bar{Q}]_1({}^{2s+1}L_J)+\gamma$ as
\begin{subequations}
\label{AQfinal}%
\begin{eqnarray}
\label{AQfinal1s0}%
\mathcal{A}_Q^{\mu\nu}({}^1S_0)
&=&
-\frac{4\sqrt{6}e_Q^2e^2}{Q^2-P^2}\,
\epsilon^{\mu\nu\alpha\beta}P_\alpha Q_\beta,
\\
\label{AQfinal3p0}%
\mathcal{A}_Q^{\mu\nu}({}^3P_0)
&=&
i\frac{8\sqrt{2}e_Q^2 e^2 (Q^2-3P^2)}{\sqrt{P^2} (Q^2-P^2)^2}\,
\left[
\frac{1}{2}(Q^2-P^2) g^{\mu\nu}
+P^\mu P^\nu
\right],
\\
\label{AQfinal3p1}%
\mathcal{A}_Q^{\mu\nu}({}^3P_1)
&=&
\frac{16\sqrt{3} e_Q^2 e^2 Q^2\epsilon^*_{H \sigma}(\lambda_H)}
     {P^2(Q^2-P^2)^2}\,
\left[
 P^\mu P_\alpha Q_\beta\epsilon^{\nu\alpha\beta\sigma} 
-P^\nu P_\alpha Q_\beta\epsilon^{\mu\alpha\beta\sigma}
-\frac{1}{2}(Q^2-P^2)P_\alpha \epsilon^{\mu\nu\alpha\sigma}
\right],
\nonumber\\
\\
\label{AQfinal3p2}%
\mathcal{A}_Q^{\mu\nu}({}^3P_2)
&=&
i\frac{16\sqrt{6}e_Q^2e^2 \sqrt{P^2}\epsilon^*_{H \alpha\beta}(\lambda_H)}
     {(Q^2-P^2)^2}\,
\bigg[
g^{\mu\nu} Q^\alpha Q^\beta
+P^\mu Q^\alpha g^{\nu\beta}
-P^\nu Q^\alpha g^{\mu\beta}
\nonumber\\
&&
+\frac{1}{2}(Q^2-P^2) g^{\mu\alpha}g^{\nu\beta}
\bigg].
\end{eqnarray}
\end{subequations}
In the limit $Q^2\to 0$, the expressions in Eq.~(\ref{AQfinal})
reduce into the effective vertices for the decay 
$[Q\bar{Q}]_1({}^{2s+1}L_J)\to\gamma\gamma$. We observe that the
vertex for ${}^3P_1$ in Eq.~(\ref{AQfinal3p1}) is proportional
to $Q^2$. Because of this reason, 
$\gamma^*\to [Q\bar{Q}]_1({}^3P_1)+\gamma$ is not forbidden
while the amplitude for $[Q\bar{Q}]_1({}^3P_1)\to \gamma \gamma$
vanishes.
\subsection{Perturbative matching
\label{sec:matching}}
The $Q\bar{Q}$ analogs $\mathcal{A}_Q^{\mu\nu}({}^{2s+1}L_J)$ 
(\ref{AQfinal}) 
for the hadronic tensors contain perturbative NRQCD matrix elements.
At the squared amplitude level, those perturbative NRQCD matrix 
elements at the leading order in $v$ are
$\langle 0|\mathcal{O}^{[Q\bar{Q}]_1({}^{2s+1}L_J)}_1%
({}^{2s+1}L_J)|0\rangle$ for the color-singlet ${}^{2s+1}L_J$ state,
where $\mathcal{O}_1$ is the local four-quark operator of 
NRQCD~\cite{Bodwin:1994jh}.
In general, the production matrix elements are related to the 
NRQCD matrix elements for the decay under vacuum-saturation 
approximation, which is valid up to corrections of relative order 
$v^4$~\cite{Bodwin:1994jh}.
For exclusive production through the electromagnetic interaction,
the $v$-leading perturbative NRQCD matrix elements 
for the ${}^1S_0$ and ${}^3P_J$ states
are simply expressed at the amplitude level as
\begin{subequations}
\label{pertME}%
\begin{eqnarray}
\label{pert1s0}%
\langle [Q\bar{Q}]_1({}^1S_0)|\psi^\dagger \chi|0\rangle
&=&2E\sqrt{2N_c},
\\
\label{pert3p0}%
\tfrac{1}{\sqrt{3}}
\langle [Q\bar{Q}]_1({}^3P_0)|
\psi^\dagger \left(-\tfrac{i}{2}
\tensor{\bm{D}}\cdot \bm{\sigma} \right)
\chi|0\rangle
&=&2E\sqrt{2N_c}|\bm{q}|,
\\
\label{pert3p1}%
\tfrac{1}{\sqrt{2}}
\langle [Q\bar{Q}]_1({}^3P_1)|
\psi^\dagger \left(-\tfrac{i}{2}
\tensor{\bm{D}}\times \bm{\sigma}\cdot 
\bm{\epsilon}_H 
\right)\chi
|0\rangle
&=&2E\sqrt{2N_c}|\bm{q}|,
\\
\label{pert3p2}%
\sum_{ij}
\langle [Q\bar{Q}]_1({}^3P_2)|
\psi^\dagger \left(-\tfrac{i}{2}
\tensor{{D}}^{(i}{\sigma}^{j)} \epsilon_H^{i j}
\right)\chi
|0\rangle
&=&2E\sqrt{2N_c}|\bm{q}|,
\end{eqnarray}
\end{subequations}
where $\psi^\dagger$ and $\chi$ are two-component Pauli spinors
creating a heavy quark and a heavy antiquark, respectively,
$\sigma^i$ is a Pauli matrix, and
$\bm{D}$ is the spatial part of the gauge covariant derivative.
Note that only a single
polarization state is projected out in the expressions
in Eqs.~(\ref{pert3p1}) and (\ref{pert3p2}). 
In Eq.~(\ref{pert3p2}), the following notation for the symmetric 
traceless component of a tensor is used:
$A^{(ij)}=\tfrac{1}{2}(A^{ij}+A^{ji})%
-\tfrac{1}{3}\delta^{ij}\sum_k A^{k k}$.
The factors of $2E$ appear on the right sides of Eq.~(\ref{pertME}),
because the state $|[Q\bar{Q}]_1({}^{2s+1}L_J)\rangle$ on the left sides
is normalized relativistically.
In Eq.~(\ref{AQP}), we have taken the derivative with respect to 
$q$. Therefore, when we read the corresponding 
short-distance coefficients, we must cancel the overall factor
$|\bm{q}|$ in Eqs.~(\ref{pert3p0}), (\ref{pert3p1}), and
(\ref{pert3p2}).
Replacement of the gauge covariant derivative $\bm{D}$ by an ordinary
derivative $\bm{\nabla}$ brings in corrections of relative order
$v^2$ in the Coulomb gauge~\cite{Bodwin:1994jh} in which
the NRQCD operators are defined.

By making use of Eqs.~(\ref{amp1s0}), (\ref{amp3pj}),
and (\ref{pertME}), we obtain the hadronic tensors:
\begin{equation}
\label{matching}%
\mathcal{A}^{\mu\nu}_{H({}^{2s+1}L_J)}
=
\sqrt{\frac{2m_H\langle \mathcal{O}_1\rangle_{L}
     }{2N_c (2E)^2 }}
\mathcal{A}_Q^{\mu\nu}({}^{2s+1}L_J).
\end{equation}
This step is equivalent to replacing the
perturbative $Q\bar{Q}$ state $|[Q\bar{Q}]_1({}^{2s+1}L_J)\rangle$
in Eq.~(\ref{AQfinal})
by the meson state $\sqrt{2m_H}|H({}^{2s+1}L_J)\rangle$,
where $m_H$ is the mass of $H$. 
At leading order in $v$, 
$m_H =2 E|_{\bm{q}=0} = 2m$.
The meson state $|H({}^{2s+1}L_J)\rangle$,
that is included in the vacuum-saturated
analog of the NRQCD decay matrix elements 
$\langle \mathcal{O}_1\rangle_{L}$ for $L=S$ or $P$,
has the nonrelativistic normalization:
\begin{subequations}
\label{nonpertME}%
\begin{eqnarray}
\label{nonpert1s0}%
\left|\langle H({}^1S_0)|\psi^\dagger \chi|0\rangle\right|^2
&=&\langle \mathcal{O}_1\rangle_{S},
\\
\label{nonpert3p0}%
\tfrac{1}{3}\left|
\langle H({}^3P_0)|
\psi^\dagger \left(-\tfrac{i}{2}
\tensor{\bm{D}}\cdot \bm{\sigma} \right)
\chi|0\rangle\right|^2
&=&\langle \mathcal{O}_1\rangle_{P},
\\
\label{nonpert3p1}%
\tfrac{1}{2}\left|
\langle H({}^3P_1)|
\psi^\dagger \left(-\tfrac{i}{2}
\tensor{\bm{D}}\times \bm{\sigma}\cdot 
\bm{\epsilon}_H 
\right)\chi
|0\rangle\right|^2
&=&\langle \mathcal{O}_1\rangle_{P},
\\
\label{nonpert3p2}%
\Big|
\sum_{ij}
\langle H({}^3P_2)|
\psi^\dagger \left(-\tfrac{i}{2}
\tensor{{D}}^{(i}{\sigma}^{j)} \epsilon_H^{i j}
\right)\chi
|0\rangle\Big|^2
&=&\langle \mathcal{O}_1\rangle_{P}.
\end{eqnarray}
\end{subequations}
The expressions in Eq.~(\ref{nonpertME}) have
errors of relative order $v^2$, which
break the heavy-quark spin symmetry.

Substituting Eq.~(\ref{AQfinal}) into Eq.~(\ref{matching}),
we obtain the hadronic tensor $\mathcal{A}_{H}^{\mu\nu}$
in Eq.~(\ref{amp0}):
\begin{subequations}
\label{AHfinal}%
\begin{eqnarray}
\label{AHfinal1s0}%
\mathcal{A}_{H({}^1S_0)}^{\mu\nu}
&=&
-\frac{4 e^2 e_Q^2}{Q^2-P^2}\,
\epsilon^{\mu\nu\alpha\beta}P_\alpha Q_\beta
\sqrt{\frac{\langle \mathcal{O}_1\rangle_{S}
     }{ m}},
\\
\label{AHfinal3p0}%
\mathcal{A}_{H({}^3P_0)}^{\mu\nu}
&=&
i\frac{4 e^2 e_Q^2(Q^2-3P^2)}{\sqrt{3}(Q^2-P^2)^2}\,
\left[
\frac{1}{2}(Q^2-P^2) g^{\mu\nu}
+P^\mu P^\nu
\right]
\sqrt{\frac{\langle \mathcal{O}_1\rangle_{P}
     }{m^3}},
\\
\label{AHfinal3p1}%
\mathcal{A}_{H({}^3P_1)}^{\mu\nu}
&=&
\frac{4 e^2 e_Q^2 Q^2\epsilon^*_{H \sigma}(\lambda_H)}
     {\sqrt{2} m(Q^2-P^2)^2}\,
\left[
 P^\mu P_\alpha Q_\beta\epsilon^{\nu\alpha\beta\sigma} 
-P^\nu P_\alpha Q_\beta\epsilon^{\mu\alpha\beta\sigma}
-\frac{1}{2}(Q^2-P^2) \epsilon^{\mu\nu\alpha\sigma}P_\alpha
\right]
\nonumber\\
&&\times
\sqrt{\frac{\langle \mathcal{O}_1\rangle_{P}
     }{ m^3}},
\\
\label{AHfinal3p2}%
\mathcal{A}_{H({}^3P_2)}^{\mu\nu}
&=&
i\frac{8e^2 e_Q^2P^2\epsilon^*_{H \alpha\beta}(\lambda_H)}
     {(Q^2-P^2)^2}\,
\left[
g^{\mu\nu} Q^\alpha Q^\beta
+P^\mu Q^\alpha g^{\nu\beta}
-P^\nu Q^\alpha g^{\mu\beta}
+\frac{1}{2}(Q^2-P^2) g^{\mu\alpha}g^{\nu\beta}
\right]
\nonumber\\
&&\times
\sqrt{\frac{\langle \mathcal{O}_1\rangle_{P}
     }{ m^3}}.
\end{eqnarray}
\end{subequations}
The hadronic tensor (\ref{AHfinal}) can be compared with
previous results for the color-octet processes
$\gamma+g^*\to [Q\bar{Q}]_8({}^{2s+1}L_J)$~\cite{Ko:1996xw}.
We find that the results in Eq.~(\ref{AHfinal}) are 
consistent with those given in Eqs.~(3.23), (3.25), (3.26),
and (3.27) of Ref.~\cite{Ko:1996xw} up to color factors.
We have also checked that Eq.~(\ref{AHfinal})
reproduces the decay widths for $\eta_Q$, $\chi_{Q0}$, and $\chi_{Q2}$
into two photons~\cite{Bodwin:1994jh} in the limit $Q^2\to 0$.
\subsection{Cross section
\label{sec:cross}}
Substituting the hadronic tensor $\mathcal{A}_{H}^{\mu\nu}$ in 
Eq.~(\ref{AHfinal}) and polarization states for the $H$ and $\gamma$
into Eq.~(\ref{amp0}), we obtain the helicity amplitude.
By squaring the helicity amplitude,
averaging over the spins of the $e^+$ and $e^-$, dividing by
the incident flux $2s$, and multiplying by the two-body phase space,
we obtain the differential cross section for each helicity state
in terms of $x=\cos \theta$, where $\theta$ is the scattering angle
of the photon at the CM frame. The resultant differential cross
section for the $e^+ e^- \to  H({}^1S_0) + \gamma$ process is 
\begin{equation}
\label{dsig1s0}%
\frac{d \sigma}{d x}[e^+ e^- \to H({}^1S_0)+ \gamma(\pm 1)] =
\frac{ \pi^2 e_Q^4 \alpha^3 r}{2 s}
\left( 1 -\frac{ m_H^2}{s} \right) ( 1+x^2) 
\frac{\langle \mathcal{O}_1 \rangle_{S}}{m^3},
\end{equation}
where $-1\le x \le 1 $ and the scaling parameter $r$ is defined by
\begin{equation}
r = \frac{4 m^2}{s}.
\end{equation}

The differential cross section of the $e^+ e^- \to H({}^3P_J) + \gamma$ process
for each helicity state is given by
\begin{equation}
\label{dsig3pj}%
\frac{d \sigma}{d x} [e^+ e^- \to H({}^3P_J)
(\lambda_H)+ \gamma(\lambda)]
=
\frac{\pi^2 e_Q^4 \alpha^3 r }
     { s \left(1-r\right)^2} F_J(\lambda_H,\lambda)
     \left( 1 -\frac{ m_H^2}{s} \right) 
     \frac{\langle \mathcal{O}_1 \rangle_{P}}{m^5},
\end{equation}
where nonvanishing entries of $F_J(\lambda_H,\lambda)$ are
\begin{subequations}
\label{F}%
\begin{eqnarray}
F_0(0,\pm 1) &=& \frac{(1-3 r)^2}{6} (1+x^2), \\
F_1(0,\pm 1) &=& 1+x^2, \\
F_1(\pm 1,\pm 1) &=& 2 r (1-x^2), \\
F_2(0,\pm 1) &=& \frac{1}{3}(1+x^2), \\
F_2(\pm 1,\pm 1) &=& 2 r (1-x^2), \\
F_2(\pm 2,\pm 1) &=& 2 r^2 (1+x^2).
\end{eqnarray}
\end{subequations}
In computing the phase space in Eqs.~(\ref{dsig1s0}) and
(\ref{dsig3pj}), we take the physical mass $m_H$ instead of the
invariant mass $2E$ of the $Q\bar{Q}$ pair.
This accounts for the factor $1-m_H^2/s$ in Eqs.~(\ref{dsig1s0}) 
and (\ref{dsig3pj}).

When the CM energy $\sqrt{s}$ is much greater than the heavy-quark
mass $m$, the asymptotic behavior of
the cross section for $e^+ e^- \to H + \gamma$ satisfies
the helicity selection rules of perturbative 
QCD~\cite{Chernyak:1980dj,Brodsky:1981kj}.
In this limit,  the scaled cross section $R(H+\gamma)$ in units of the
cross section for $\mu^+\mu^-$ approaches the asymptotic 
form~\cite{Braaten:2002fi}
\begin{equation}
\label{R}%
R[H(\lambda_H)+\gamma] \sim 
\alpha v^{3+2n(L)} r^{1+|\lambda_H|},
\end{equation}
where $n(S)=0$ and $n(P)=1$. The factor $v^{3+2n(L)}$ in Eq.~(\ref{R})
arises from the wave function of the $H$.
One power of $r$ in Eq.~(\ref{R}) stems from the large momentum transfer
which is required for the heavy-quark pair to form the heavy quarkonium
with small relative momentum and the other powers arise from the 
helicity selection rules~\cite{Braaten:2002fi}.
In the limit $r\to 0$, the differential cross sections 
in Eqs.~(\ref{dsig1s0}) and (\ref{dsig3pj}) 
satisfy the asymptotic form (\ref{R}).
For the exclusive charmonium production associated with a photon
at the $B$ factories, the ratio $r\approx 0.07$ is small for $m_c=1.4$~GeV.
However, for the case of a bottomonium,
$r\approx 0.76$ is of order 1 for $m_b=4.6$~GeV. Therefore, the asymptotic
behavior (\ref{R}) is approximately satisfied only for the exclusive
charmonium plus $\gamma$ production at the $B$ factories.

Integrating the differential cross section given in Eq.~(\ref{dsig1s0}) 
over $x$ and summing  over the photon helicities,
we find the total cross section for the process
$e^+ e^- \to H(^1S_0) + \gamma$:
\begin{equation}
\label{sig1s0}%
\sigma[e^+ e^- \to H({}^1S_0)+\gamma] =
\frac{8 \pi^2 e_Q^4 \alpha^3 r}{3  s}
\left(1-\frac{m_H^2}{s}\right)
\frac{\langle \mathcal{O}_1 \rangle_S}{m^3}.
\end{equation}
Integrating Eq.~(\ref{dsig3pj}) over $x$ and summing over the 
helicities of the photon and quarkonium, we obtain the total
cross sections for $e^+ e^- \to H({}^3P_J) +\gamma$:
\begin{subequations}
\label{sig3pj}%
\begin{eqnarray}
\sigma [e^+ e^- \to H({}^3P_0)+ \gamma] &=&
\frac{8\pi^2 e_Q^4 \alpha^3 r \left( 1-3 r\right)^2 }
     {9 s \left(1-r\right)^2}
\left( 1 -\frac{m_H^2}{s}\right) 
\frac{\langle \mathcal{O}_1 \rangle_{P}}{m^5},
\\
\sigma[e^+ e^- \to H({}^3P_1) + \gamma] &=&
\frac{16 \pi^2 e_Q^4 \alpha^3 r (1+r) }
     {3 s \left(1-r\right)^2}
\left( 1-\frac{m_H^2}{s} \right)
\frac{\langle \mathcal{O}_1 \rangle_{P}}{m^5},
\\
\sigma[e^+ e^- \to H({}^3P_2)+\gamma] &=&
\frac{16 \pi^2 e_Q^4 \alpha^3 r(1+3 r+6 r^2) }
     {9 s \left(1-r\right)^2}
\left( 1-\frac{m_H^2}{s} \right)
\frac{\langle \mathcal{O}_1 \rangle_{P}}{m^5},
\end{eqnarray}
\end{subequations}
where we have used the spin-1 and spin-2 polarization
tensors for ${}^3P_1$ and ${}^3P_2$ states summed over
polarization states:
\begin{subequations}
\begin{eqnarray}
\sum_{\lambda_H} \epsilon_H^\mu (\lambda_H)
\epsilon_H^{\ast\,\alpha} (\lambda_H)
&=&
I^{\mu\alpha},
\\
\sum_{\lambda_H} \epsilon_H^{\mu\nu}(\lambda_H) 
\epsilon_H^{\ast\,\alpha\beta}(\lambda_H)
&=&  
\frac{1}{2}(I^{\mu\alpha}I^{\nu\beta}+I^{\mu\beta}I^{\nu\alpha})
-\frac{1}{3}I^{\mu\nu}I^{\alpha\beta}.
\end{eqnarray}
\end{subequations}

\section{Parameters for the numerical analysis
\label{sec:input}}
In order to predict the cross sections for $e^+e^-\to \eta_Q+\gamma$ 
and $\chi_{QJ}+\gamma$ at the
$B$ factories by using Eqs.~(\ref{sig1s0}) and (\ref{sig3pj}),
we need to determine the numerical values for the NRQCD
matrix elements $\langle \mathcal{O}_1\rangle_{L}$ and input
parameters such as $m$, $m_H$, $\alpha$, and $\alpha_s$.
In this section, we list the numerical values for those quantities
that we use in this work.
\subsection{\label{param}Input parameters}
The short-distance coefficients for the cross sections in 
Eqs.~(\ref{sig1s0}) and (\ref{sig3pj}) depend on the heavy-quark mass
$m$. We take the one-loop pole mass for the
heavy-quark mass $m$: $m_c=1.4 \pm 0.2 $~GeV for the charm quark and 
$m_b= 4.6 \pm 0.1$~GeV for the bottom quark, respectively.
As we mentioned earlier, we use the physical quarkonium mass for $m_H$
in the phase-space factor $1-m_H^2/s$ in Eqs.~(\ref{sig1s0}) and 
(\ref{sig3pj}), where $m_{\eta_c}=2.9804$~GeV, $m_{\eta_c(2S)}=3.638$~GeV,
$m_{\chi_{c0}} = 3.41475$~GeV, $m_{\chi_{c1}}=3.51066$~GeV,
$m_{\chi_{c2}} = 3.55620$~GeV, $m_{\chi_{b0}}=9.585944$~GeV,
$m_{\chi_{b1}} = 9.89278$~GeV, 
and $m_{\chi_{b2}}=9.991221$~GeV~\cite{Yao:2006px}.
For the $\eta_b$ mass, we take  
$m_{\eta_b}=9.3889$ GeV~\cite{:2008vj}.\footnote{
Very recently, the BABAR Collaboration observed the
$\eta_b$ resonance in the photon energy spectrum in radiative
 $\Upsilon(3S)$ decay~\cite{:2008vj}.
}

The cross sections in Eqs.~(\ref{sig1s0}) and (\ref{sig3pj}) are
proportional to $\alpha^3$. For the QED couplings $\alpha^2$ for the
virtual photon, we use the running coupling constant~\cite{Erler:1998sy}
$\alpha(\mu=\sqrt{s})= 1/131$ at $\sqrt{s}=10.58$~GeV.
The QED coupling for the real photon is chosen to be 
$\alpha(\mu= z) \approx 1/132$, where $z = [(s - 2 m^2)/2]^{1/2}$ is 
the invariant mass of the virtual heavy quark:
$z \approx 7.3$~GeV and $5.9$~GeV for the charmonium and the bottomonium,
respectively.

\subsection{\label{ME}Long-distance matrix elements}
\subsubsection{\label{MES}
NRQCD matrix elements ${\langle \mathcal{O}_1\rangle_S}$}
The color-singlet NRQCD matrix elements for the $S$-wave spin-singlet 
charmonia, $\eta_c$ and $\eta_c(2S)$, can be determined from the
two-photon decay rates. However, up to errors of relative order $v^2$
that arise from the heavy-quark-spin-symmetry breaking, one can determine 
the values also by making use of the leptonic decay rates of the 
spin-triplet counterparts, which are far more accurately measured.
Among various theoretical attempts to determine the matrix elements
for the $S$-wave spin-singlet charmonia, we employ a recently
developed method given in
Refs.~\cite{Bodwin:2007fz,Bodwin:2006dn,Chung:2008sm}.
This method resums a class of relativistic corrections to the color-singlet
contributions to the electromagnetic decay rates of both
spin-singlet and triplet $S$-wave charmonia to all orders in $v$. 
The color-singlet NRQCD matrix element $\langle \mathcal{O}_1\rangle_S$
for the $\eta_c$ is taken to be
$0.437^{+0.111}_{-0.105}~\textrm{GeV}^3$~\cite{Bodwin:2007fz}.

For the radially excited $S$-wave spin-singlet state $\eta_c(2S)$, 
the NRQCD matrix element determined from the measured decay width for
$\eta_c(2S)\to \gamma\gamma$ differs significantly 
by about $4.6 \sigma$ from that for $\psi(2S)$ determined from 
$\psi(2S)\to e^+e^-$.
The difference is greater than the errors of neglecting spin-symmetry
breaking effects.
We suspect that the  main reason for the discrepancy is originated
from the assumption~\cite{Athar:2004dn}:
\begin{equation}
\label{assumption}%
\textrm{Br}[\eta_c(2S)\to K_S^0 K^\pm \pi^\mp]=
\textrm{Br}[\eta_c\to K_S^0 K^\pm \pi^\mp].
\end{equation}
This assumption was imposed on the indirect determination of
the partial width for $\eta_c(2S)\to \gamma\gamma$ from the
measurement for $\gamma^\ast \gamma^\ast \to \eta_c(2S)%
\to K_S^0 K^\pm \pi^\mp$~\cite{Athar:2004dn}.
In this work, instead of using this indirect determination
of $\Gamma[\eta_c(2S)\to \gamma\gamma]$ in 
Ref.~\cite{Athar:2004dn}, we assume that
the NRQCD matrix element determined from
$\Gamma[\eta_c(2S)\to \gamma \gamma]$ must be the same as
that determined from $\Gamma[\psi(2S)\to e^+e^-]$ within
errors of relative order $v^2$, which breaks the heavy-quark
spin symmetry. Our assumption results in the prediction for
the two-photon width of $\eta_c(2S)$:
$\Gamma[\eta_c(2S)\to \gamma\gamma]=2.92$~keV.
Direct measurement of the two-photon width of $\eta_c(2S)$
may provide one with a stringent test of our assumption.
We will present
a detailed analysis regarding this point in a separate publication.
Basic strategy for determining the matrix element is similar to
that in Ref.~\cite{Bodwin:2007fz}.
Disregarding the assumption (\ref{assumption}) and
making use of the matrix element obtained from
$\Gamma[\psi(2S)\to e^+ e^-]$ based on the heavy-quark spin symmetry
up to corrections of relative order $v^2$, we find that
$\langle \mathcal{O}_1 \rangle_S = 
0.274^{+0.042}_{-0.036}~\textrm{GeV}^3~\textrm{for}~S=\eta_c(2S)$.

The NRQCD matrix element $\langle \mathcal{O}_1 \rangle_{S}$ 
for the $\eta_b$ cannot
be determined from $\eta_b\to\gamma\gamma$ because the measurement is
not available, yet. Instead, we quote the value for the matrix element
$\langle \mathcal{O}_1 \rangle_{\Upsilon(1S)}$ given in Eq.~(23a) of
Ref.~\cite{Kang:2007uv}, which is determined from 
$\Upsilon(1S)\to e^+e^-$. We assume that the heavy-quark spin symmetry
is relatively well satisfied because
$v^2\sim 0.1$ is small for the bottomonium. The numerical value is 
$\langle \mathcal{O}_1 \rangle_S =%
3.069^{+0.207}_{-0.190}~\textrm{GeV}^3$ for $S=\eta_b$~\cite{Kang:2007uv}.
\subsubsection{\label{sec:MEP}
NRQCD matrix elements ${\langle \mathcal{O}_1\rangle_P}$}
The generalization of the resummation of relativistic 
corrections~\cite{Bodwin:2006dn} to the $P$-wave
quarkonium has not been developed because of the complication due to
the color-octet contributions. Therefore, for the spin-triplet $P$-wave 
charmonium $\chi_{cJ}$, we determine the NRQCD matrix element 
$\langle \mathcal{O}_1\rangle_P$ by comparing the $v$-leading NRQCD
factorization formulas for the photonic decay widths~\cite{Bodwin:1994jh}
\begin{subequations}
\begin{eqnarray}
\label{gamma0aa}%
\Gamma(\chi_{c0}\to \gamma\gamma) &=&
6\pi e_c^4 \alpha^2 m \left[ 1+\frac{(3\pi^2-28)\alpha_s}{18\pi}\right]^2
\frac{\langle \mathcal{O}_1 \rangle_{P}}{m^5}, 
\\
\label{gamma2aa}%
\Gamma(\chi_{c2}\to \gamma\gamma) &=&
\frac{8\pi e_c^4 \alpha^2 m}{5} 
\left( 
1-\frac{8\alpha_s}{3\pi}
\right)^2
\frac{\langle \mathcal{O}_1 \rangle_{P}}{m^5} 
\end{eqnarray}
\end{subequations}
with corresponding measured values.
The decay of $\chi_{c1}$ into $\gamma\gamma$ 
is absent due to Yang's theorem~\cite{Yang:1950rg}.

Substituting the numerical values 
$\textrm{Br}(\chi_{c0}\to \gamma\gamma) = (2.76\pm 0.33)\times 
10^{-4}$~\cite{Yao:2006px},
the total decay width $\Gamma_{\chi_{c0}}=10.4\pm 0.7$~MeV~\cite{Yao:2006px}, 
$\alpha_s (\mu=m_{\chi_{cJ}}/2)=0.32$, and 
$\alpha(\mu=m_{\chi_{cJ}}/2)=1/133$~\cite{Erler:1998sy}
into Eq.~(\ref{gamma0aa}),
we obtain 
$\langle \mathcal{O}_1 \rangle_{P}=%
\langle \mathcal{O}_1 \rangle_{\chi_{c0}}^{\gamma\gamma}%
= 0.051 \pm 0.010$~GeV${}^5$ for $P=\chi_{c0}$, 
where the superscript $\gamma\gamma$
indicates that the value is fit to the two-photon width.
Taking into account $\textrm{Br}(\chi_{c2}\to \gamma\gamma) = (2.58 \pm 0.19)
\times 10^{-4}$~\cite{Yao:2006px}
and $\Gamma_{\chi_{c2}} = 2.05\pm 0.12$~MeV~\cite{Yao:2006px}, we obtain
the matrix element 
$\langle \mathcal{O}_1 \rangle_P =
\langle \mathcal{O}_1 \rangle_{\chi_{c2}}^{\gamma\gamma}
= 0.068 \pm 0.009$~GeV${}^5$ for $P=\chi_{c2}$.
Here, the uncertainties from the heavy-quark mass are not included in both
$\langle \mathcal{O}_1 \rangle_{\chi_{c0}}^{\gamma\gamma}$ and
$\langle \mathcal{O}_1 \rangle_{\chi_{c2}}^{\gamma\gamma}$.
In Ref.~\cite{Braaten:2002fi}, 
the value $\langle \mathcal{O}_1 \rangle_{\chi_{c2}} 
= 0.053 \pm 0.009$~GeV${}^5$ was obtained by making use of a similar method.
The main reason for the difference is
the change in the measured branching fraction 
of the $\chi_{c2}$ decay into $\gamma\gamma$
from $(2.19\pm 0.33)\times 10^{-4}$~\cite{Hagiwara:2002fs} 
to $(2.58 \pm 0.19)\times 10^{-4}$~\cite{Yao:2006px}.
We take the weighted average over the spin states of 
$\chi_{c0}$ and $\chi_{c2}$ to determine
$\langle  \mathcal{O}_1 \rangle_{P} =%
0.060^{+0.043}_{-0.029}~\textrm{GeV}^5\textrm{ for }P=\chi_{cJ}$,
where the heavy-quark spin symmetry is imposed
and the uncertainty from the heavy-quark mass is considered together with
those from the experimental values.

In the cases of the spin-triplet $P$-wave bottomonium states $\chi_{bJ}(1P)$,
the measured values for the two-photon decay widths are not available.
Therefore, we use the first derivative of the wave function at the origin
that has been determined by using 
the Buchm\"uller-Tye potential~\cite{Eichten:1995ch}. 
The resultant value for the NRQCD matrix element is
$\langle  \mathcal{O}_1 \rangle_{P} =%
2.03~\textrm{GeV}^5\textrm{ for }P=\chi_{bJ}$~\cite{Bodwin:2007zf}.
\subsubsection{Summary of NRQCD matrix elements}
In Table~\ref{tableME}, we tabulate the color-singlet NRQCD matrix
elements for $\eta_c$, $\eta_c(2S)$, $\eta_b$, $\chi_{cJ}$,
and $\chi_{bJ}$. 
As has been described in the text, there are input parameters
such as $m$ that have been used to determine the NRQCD matrix elements.
When we compute the short-distance coefficients for the cross section
for $e^+ e^- \to H+\gamma$, such input parameters must be consistent with 
the values that have been used to determine the NRQCD matrix 
elements~\cite{Bodwin:2007fz}. In order to take into account such
correlations, we present sources of the uncertainties
for each NRQCD matrix element in Table~\ref{tableME}.
The first and the second lines after headings contain uncertainties
arising from the heavy-quark mass $m$ and from sources described
in each reference except for $m$, respectively.
The sum of the two uncertainties is obtained by adding them in quadrature 
as shown in the last row of Table~\ref{tableME}.

\begin{table}[t]
\caption{\label{tableME}%
NRQCD matrix elements $\langle \mathcal{O}_1 \rangle_H$ 
for $H(^1S_0)$ in units of GeV$^3$ and 
for $H(^3P_J)$ in units of GeV$^5$. 
The first and the second lines after headings contain uncertainties
arising from the heavy-quark mass $m$ and from other sources described
in the text.
Uncertainties in the last line are obtained by adding the two uncertainties
in quadrature.
}
\begin{ruledtabular}
\begin{tabular}{l|ccccc}
Sources of errors
$\backslash$ $H$ 
 & $\eta_c$~\cite{Bodwin:2007fz}
 & $\eta_c(2S)$
 & $\eta_b$~\cite{Bodwin:2007zf}
 & $\chi_{cJ}$
 & $\chi_{bJ}$~\cite{Bodwin:2007zf,Eichten:1995ch}
\\
 & (GeV$^3$)
 & (GeV$^3$)
 & (GeV$^3$)
 & (GeV$^5$)
 & (GeV$^5$)
\\
\hline
$\Delta m$ &
  $0.437^{+0.033}_{-0.025}$ &
  $0.274^{+0.013}_{-0.010}$ &
  $3.069^{+0.000}_{-0.000}$ &
  \!\!\!\!\!\!\!$0.060^{+0.043}_{-0.028}$ &
  $2.03$
\\
others &
  $0.437^{+0.106}_{-0.102}$ &
  $0.274^{+0.040}_{-0.035}$ &
  $3.069^{+0.207}_{-0.190}$ &
  $0.060 \pm 0.007$ &
  $2.03$
\\
\hline
total &
  $0.437^{+0.111}_{-0.105}$ &
  $0.274^{+0.042}_{-0.036}$ &
  $3.069^{+0.207}_{-0.190}$ &
  \!\!\!\!\!\!\!$0.060^{+0.043}_{-0.029}$ &
  $2.03$
\end{tabular}
\end{ruledtabular}
\end{table}
\section{\label{sec:num}
Predictions for the $\bm{B}$ factories}
In this section, we provide the predictions for the cross sections for
$e^+e^-\to H+\gamma$ 
at the $B$ factories. We also discuss the uncertainties and
phenomenological implications of the results.

Our predictions for the total cross section $\sigma$ for 
$e^+e^-\to H+\gamma$ at the CM energy $\sqrt{s}=10.58$~GeV are given
in the first column of 
Table~\ref{tablexsec}. The values have been obtained by substituting
the input parameters listed in Sec.~\ref{sec:input} into the NRQCD
factorization formulas in Eqs.~(\ref{sig1s0}) and (\ref{sig3pj}).
In the second column, we provide the cross section 
$\sigma_{\textrm{cut}}$ in which we have imposed the cut $|\cos\theta|<0.8$ 
for the scattering angle $\theta$ of the photon at the CM frame.
$\sigma_{\textrm{cut}}$ is about $70$\%--$80$\% of $\sigma$ in any case.
In the last column in Table~\ref{tablexsec}, we list 
the energy $E_\gamma=(s-m_H^2)/(2\sqrt{s})$ of the photon 
emitted in $e^+e^-\to H+\gamma$ at 
the CM frame, which may provide an efficient trigger for the signals.

\begin{table}[t]
\caption{\label{tablexsec}%
The total cross section $\sigma$ in units of fb and the photon
energy $E_\gamma$ at the CM frame in units of GeV in $e^+e^- \to H+\gamma$
depending on charge-conjugation parity $C=+1$ quarkonium $H$.
The subscript ``cut'' represents the cut $|\cos\theta|<0.8$ 
for the scattering angle $\theta$ of the photon at the CM frame. 
}
\begin{ruledtabular}
\begin{tabular}{l|ccc}
$H$ & $\sigma$~(fb)& $\sigma_{\textrm{cut}}$~(fb)
& $E_\gamma$~(GeV)
\\
\hline
$\eta_c$ & $82.0^{+21.4}_{-19.8} $ 
         & $59.7^{+15.6}_{-14.4} $ 
         & $4.87$ 
\\
$\eta_c(2S)$ & $49.2^{+9.4}_{-7.4} $ 
             & $35.8^{+6.8}_{-5.4} $ 
             & $4.66$ 
\\
$\eta_b$ & $\phantom{0}2.5^{+0.2}_{-0.2} $ 
         & $\phantom{0}1.8^{+0.1}_{-0.1} $ 
         & $1.12$ 
\\
\hline
$\chi_{c0}$ & $\phantom{0}1.3^{+0.2}_{-0.2} $ 
            & $\phantom{0}1.0^{+0.1}_{-0.1} $ 
            & $4.74$ 
\\
$\chi_{c1}$ & $13.7^{+3.4}_{-3.1} $ 
            & $10.2^{+2.6}_{-2.3} $ 
            & $4.71$ 
\\
$\chi_{c2}$ & $\phantom{0}5.3^{+1.6}_{-1.3} $ 
            & $\phantom{0}4.0^{+1.3}_{-1.0} $ 
            & $4.69$ 
\\
\hline
$\chi_{b0}$ & $\phantom{0}0.6^{+0.3}_{-0.2} $ 
            & $\phantom{0}0.4^{+0.2}_{-0.1} $ 
            & $0.95$ 
\\
$\chi_{b1}$ & $\phantom{0}2.8^{+0.8}_{-0.5} $ 
            & $\phantom{0}2.3^{+0.6}_{-0.4} $ 
            & $0.66$ 
\\
$\chi_{b2}$ & $\phantom{0}3.0^{+1.0}_{-0.7} $ 
            & $\phantom{0}2.4^{+0.8}_{-0.5} $ 
            & $0.57$ 
\\
\end{tabular}
\end{ruledtabular}
\end{table}
\subsection{Uncertainties}
The error bars of the cross sections $\sigma$ and $\sigma_{\textrm{cut}}$
listed in Table~\ref{tablexsec} contain the following uncertainties.
A dominant source of the uncertainties is the heavy-quark mass $m$,
which affects both the NRQCD matrix elements in Table~\ref{tableME}
and the short-distance coefficients appearing in the NRQCD factorization
formulas in Eqs.~(\ref{sig1s0}) and (\ref{sig3pj}).
To avoid double counting of those errors, we have considered the
correlation of the uncertainties from $m$. 
Other errors in $\langle \mathcal{O}_1 \rangle_S$ 
include a theoretical uncertainty of relative order $v^2$ 
from the leading-potential approximation, 
the uncertainty of the string tension, which is 
a parameter of the Cornell potential model,
the uncertainty reflecting the uncalculated next-to-next-to-leading-order 
corrections in $\alpha_s$ to the electromagnetic decay width of the 
quarkonium, and the uncertainty of the measured decay width.
For more details about the error analysis,
we refer the reader to Refs.~\cite{Bodwin:2007fz}.
The error bars of the 
NRQCD matrix elements $\langle \mathcal{O}_1 \rangle_P$
for the $\chi_{cJ}$ in Table~\ref{tableME} include the
uncertainties from the heavy-quark mass
$m$ and the measured widths for the two-photon decays of 
the $\chi_{c0}$ and $\chi_{c2}$. 
The NRQCD matrix elements $\langle \mathcal{O}_1 \rangle_P$
for the $\chi_{bJ}$ are quoted from 
Refs.~\cite{Bodwin:2007zf,Eichten:1995ch}, 
where uncertainties are not included~\cite{Bodwin:2007zf}.
As a result, except for the uncertainties from the heavy-quark mass
$m$ appearing in the short-distance coefficients,
no additional uncertainties are included in the error bars
of the cross section for $H=\chi_{bJ}$.
The error bars of $\sigma$ and $\sigma_{\textrm{cut}}$
in Table~\ref{tablexsec} have been obtained by
adding all the uncertainties listed above in quadrature.

Our predictions in Table~\ref{tablexsec} may suffer from 
QCD and relativistic corrections that have not been calculated, yet.
We guess that the corrections may not be
as large as those observed in the two-charmonium production from 
$e^+ e^-$ annihilation~\cite{Zhang:2005cha,Bodwin:2006ke,Bodwin:2007ga,
He:2007te}, where all such corrections are
aligned to make a significant enhancement of the prediction for 
the cross section.
\subsection{Exclusive $\bm{\eta_c(1S,2S)+\gamma}$ production}
As we have shown in Table~\ref{tablexsec}, the cross sections for 
the $S$-wave spin-singlet charmonia $H=\eta_c$ and $\eta_c(2S)$ produced 
exclusively with a photon are about $80$ fb and about $50$ fb,
respectively.
These values are significantly greater than those for $J/\psi+\eta_c$ and 
$J/\psi+\eta_c(2S)$ measured at the $B$ factories.
We have predicted that $\Gamma[\eta_c(2S)\to\gamma\gamma]=2.92$~keV
by replacing  the assumption (\ref{assumption}) that was imposed on
the indirect determination of  $\Gamma[\eta_c(2S)\to\gamma\gamma]$ in
Refs.~\cite{Asner:2003wv,Belleetac2S}
with an alternative assumption that the approximate heavy-quark spin
symmetry still holds
between the NRQCD matrix elements for $\eta_c(2S)$ and $\psi(2S)$.
The measurement of the cross section for
$e^+ e^- \to \eta_c(2S) + \gamma$
is particularly interesting because the measurement may provide an
independent test of our argument presented in Sec.~\ref{MES}.

\subsection{Exclusive $\bm{\eta_b+\gamma}$ production}
Our prediction of  the cross section for $H=\eta_b$
 at the $B$ factories is
about $2.5$ fb. By taking into account the 
integrated luminosities available at the present $B$ factories,
we expect that over three thousands of $e^+e^-\to \eta_b+\gamma$ events can be
produced. This number exceeds the estimated number of events at LEP II
through $\gamma^\ast \gamma^\ast \to \eta_b$ by about an order of 
magnitude~\cite{Heister:2002if,Levtchenko:2004ku,Sokolov:2004kv}.
Therefore, the exclusive process $e^+ e^- \to \eta_b + \gamma$
could be an alternative probe to the $\eta_b$ meson
as long as the background can be removed significantly.
\subsection{Exclusive $\bm{\chi_{QJ}+\gamma}$ production}
As shown in Table~\ref{tablexsec},
the cross section for $H=\chi_{QJ}$ ranges from about 1 to 5 fb except for
the case of $\chi_{c1}$, which has a particularly large value of
about $14$ fb.
The reason is that  
the overall coefficient for the longitudinal $\chi_{Q1}$ is greater
than those for the other two states as shown in Eq.~(\ref{F}).\footnote{
According to the hadron helicity selection rules,
the longitudinal cross section for $\chi_{QJ}$ dominates as $r\to 0$.
}
This contrasts to the $P$-wave decay into $\gamma\gamma$, where 
$\chi_{Q1}\to \gamma\gamma$ is forbidden by Yang's theorem~\cite{Yang:1950rg}.
A similar feature can be found in the inclusive charm production
in $\chi_{bJ}$ decay~\cite{Bodwin:2007zf}: the decay rate of the $\chi_{b1}$ 
into $g^* g$ is comparable to those of the $\chi_{b0}$ and 
the $\chi_{b2}$, especially in the limit of 
the large invariant mass of the $g^\ast$.
Unlike the $P$-wave spin-triplet charmonium, the cross sections for
$\chi_{b1}$ and $\chi_{b2}$ are comparable and that for $\chi_{b0}$
is much suppressed.
\subsection{Backgrounds
\label{background}}
The exclusive process $e^+ e^- \to H +\gamma$ for $C=+1$ heavy quarkonium
$H$ suffers from large contamination from the background 
$e^+e^-\to X+\gamma$. In this section, we provide a rough estimate
on the background.

We assume that the dominant subprocesses of the background
$e^+ e^- \to X+\gamma$ are $e^+ e^- \to q\bar{q}+\gamma$,
where $q=u$, $d$, $s$, and $c$. These processes have collinear divergences
due to the photon radiation from the initial leptons and final massless
quarks. In order to avoid such complications, we take the following
cuts: $|\cos\theta|<0.8$, $\cos\theta_{q\gamma}<0.8$,
and $\cos\theta_{\bar{q}\gamma}<0.8$, where $\theta$ is the scattering
angle of the photon and $\theta_{i\gamma}$ is the relative angle between
the photon and a quark or an antiquark $i$ in the CM frame.
We also restrict the energy range of the photon by requiring the
recoil mass $m_X$ to satisfy $|m_X-m_H|<50$~MeV.

By imposing the constraints listed above, we carry out the numerical
calculation for the background cross section $\sigma_{\textrm{BG}}(H)$
for each $C=+1$ heavy quarkonium 
$H$ by making use of CompHEP~\cite{Boos:2004kh}.
The results are
$\sigma_{\textrm{BG}}(\textrm{charmonium})\approx 0.5$~pb and
$\sigma_{\textrm{BG}}(\textrm{bottomonium})\approx 2.4$--$4.5$~pb.
The background cross sections $\sigma_{\textrm{BG}}(H)$ are
much greater than  the signal cross sections
$\sigma_\textrm{cut}$ in Table~\ref{tablexsec}.
However, the  high integrated luminosities $\mathcal{L}$
of order $1$ ab$^{-1}$ of the present $B$ factories
might allow one to measure the exclusive $C=+1$ quarkonium
production associated with a photon. 
We can compute the signal significance $S(H)$ for each heavy quarkonium,
where $S(H)$ is defined by
\begin{equation}
S(H)=
\frac{\sigma_\textrm{cut}(H+\gamma)\mathcal{L}}
     {\sqrt{\sigma_{\textrm{BG}}(H) \mathcal{L} }}.
\end{equation}
We find that $S$-wave charmonium states have particularly large
values: $S(\eta_c)\approx 81$ and $S[\eta_c(2S)]\approx 47$.
In the cases of $P$-wave charmonium states,
$S(\chi_{c0})\approx 1$, $S(\chi_{c1})\approx 13$, and
$S(\chi_{c2})\approx 5$. And $S(H)$ is less than 1 for
$\eta_b$ and $\chi_{bJ}$ states. 

Our background study given above is only a rough estimate.
In the cases of bottomonium states, there must be obstacles in
detecting photons with energies about $1$ GeV or less because of the
large contamination from the decays of light hadrons~\cite{Uehara}. 
However, the exclusive charmonium production associated with a photon 
may have significantly less additional backgrounds unlike the 
bottomonium states because the radiating photon is sufficiently hard.
We also note that there may be another background arising from the
feeddown from the $e^+ e^- \to H(^3S_1)+\gamma$ process for
$H(^3S_1)= J/\psi$, $\psi(2S)$, or $\Upsilon(nS)$, whose cross sections
are of order pb. Especially, the measurement of
the $\chi_{QJ}+\gamma$ production may be affected by such feeddown.
Reduction of such backgrounds may be subject to detector resolution.

\section{Summary
\label{sec:summary}}
We have calculated the cross section for the 
$C=+1$ heavy quarkonium $H$ produced exclusively with a photon from
$e^+e^-$ annihilation into a virtual photon
at leading order in $\alpha_s$ and $v$.
The NRQCD factorization formulas for the differential distributions 
with respect to the scattering angle of the photon and the total cross sections
for $H=\eta_{Q}$ and $\chi_{QJ}$ have been obtained 
in Eqs.~(\ref{dsig1s0}), (\ref{dsig3pj}), (\ref{sig1s0}), 
and (\ref{sig3pj}), respectively.
Our predictions for the cross sections at $\sqrt{s}=10.58$~GeV
are tabulated in Table~\ref{tablexsec}.
The predicted cross sections for the $S$-wave spin-singlet states are 
about $80$~fb, $50$~fb, and $3$ fb for $\eta_c$, $\eta_c(2S)$,
and $\eta_b$, respectively. In the cases of the $P$-wave spin-triplet
charmonia, the cross section for $\chi_{c1}$ is about
14 fb; that is significantly greater than those for $\chi_{c0}$ and
$\chi_{c2}$. The cross sections for $\chi_{b1}$ and $\chi_{b2}$ are
about 3 fb and that for $\chi_{b0}$ is suppressed.
Under the cut $|\cos \theta| < 0.8$ for the scattering angle $\theta$
of the photon, the cross sections decrease by about 20\%--30 \%. 

The exclusive process $e^+e^-\to H+\gamma$ may suffer from a
large background from $e^+e^-\to X+\gamma$.
A rough estimate on the background cross section has been made by
considering only the $e^+ e^- \to q\bar{q}+\gamma$ contributions
at leading order in $\alpha_s$.
In computing the background cross section,
we have required $|m_X-m_H|<50$~MeV for the recoil
mass $m_X$ and
imposed the angular cuts $|\cos\theta|<0.8$,
$\cos\theta_{q\gamma}<0.8$, and $\cos\theta_{\bar{q}\gamma}<0.8$.
The resultant background cross sections are about $0.5$ pb for the
charmonia and about several pb for the bottomonia. The signal
significances for the $\eta_c$, $\eta_c(2S)$, $\chi_{c1}$,
and $\chi_{c2}$ cases are greater than $6$. 
However, those for the 
$\eta_b$, $\chi_{c0}$, and $\chi_{bJ}$ 
are about $1$  at the present $B$ factories.

Very recently, the BABAR Collaboration 
observed the $\eta_b$ resonance in the photon energy spectrum in radiative
 $\Upsilon(3S)$ decay~\cite{:2008vj}.
The exclusive process $e^+e^-\to \eta_b+\gamma$ that we have considered  
could be an alternative probe to the $\eta_b$ meson
as long as the backgrounds are significantly removed.

As we have remarked earlier, the authors of 
Refs.~\cite{Asner:2003wv} and \cite{Belleetac2S} made an
assumption $\textrm{Br}[\eta_c(2S)\to K_S^0 K^\pm \pi^\mp]=%
\textrm{Br}[\eta_c\to K_S^0 K^\pm \pi^\mp]$ in determining the
$\Gamma[\eta_c(2S)\to \gamma\gamma]=1.3\pm 0.6$~keV~\cite{Asner:2003wv} 
and $0.59\pm 0.19$~keV~\cite{Belleetac2S}, respectively.
This assumption seems to be the origin of
inconsistency of the NRQCD matrix element for
$\eta_c(2S)$, 
which was fit to $\Gamma[\eta_c(2S)\to \gamma\gamma]$~\cite{Asner:2003wv,%
Belleetac2S},
in comparison with that for $\psi(2S)$.
As an alternative choice, we assume that the approximate heavy-quark spin
symmetry holds between the NRQCD matrix elements for the $\eta_c(2S)$
and $\psi(2S)$ to predict $\Gamma[\eta_c(2S)\to \gamma\gamma]=2.92$ keV.
The measurement of the cross sections for
$e^+ e^- \to \eta_c+\gamma$ and $e^+ e^- \to \eta_c(2S) +\gamma$ may provide a
stringent independent constraint to test our argument.

\begin{acknowledgments}
We thank Sadaharu~Uehara and Eunil~Won for critical comments
and useful discussions.
The work of HSC was supported by the BK21 program.
The work of JL was supported by the Korea Science and Engineering Foundation
(KOSEF) funded by the Korea government (MEST) 
under Grant No. R01-2008-000-10378-0.
The work of CY was supported by the
Korea Research Foundation Grant funded by the Korean Government
(MOEHRD, Basic Research Promotion Fund) (KRF-2006-311-C00020).
\end{acknowledgments}

\end{document}